\documentstyle[fleqn,sprocl]{article}

\newcommand{\LQ}{\Lambda_{QCD}}
\newcommand{\be}{\begin{equation}}
\newcommand{\ee}{\end{equation}}
\newcommand{\bea}{\begin{eqnarray}}
\newcommand{\eea}{\end{eqnarray}}
\newcommand{\aQ}{\alpha_s(Q^2)}
\newcommand{\ak}{\alpha_s(k^2_{\perp})}
\newcommand{\am}{\alpha_s(\mu^2)}
\newcommand{\ma}{monopole-antimonopole}

\newcommand{\diff}{\partial}

\begin{document}
\title{ Uncovering thick and thin strings via renormalons.}

\author{V.I. Zakharov}\address{Max-Planck Institut f\"ur Physik,
        80805 Munich, Germany.}

\maketitle

\begin{abstract}
The talk is about
the power corrections in QCD. 
Renormalons, both infrared and ultraviolet, provide with a kind of a 
kinematical framework
to fix exponents of the leading power corrections to 
various observables. 
Any viable dynamical framework is to reproduce this pattern of the power 
corrections. In this mini-review, 
we emphasize that a simple dynamical framework 
satisfying this requirement may well be provided
by the strings inherent to the Abelian Higgs model. This model
comes naturally into consideration within the $U(1)$ projection of
QCD which makes explicit the dual-superconductor model of the confinement.
At large distances (compared to $\LQ^{-1}$) 
there are Abrikosov-Nielsen-Olesen strings which
develop a characteristic
non-perturbative
transverse size of order $\LQ^{-1}$ and match in this way
the infrared renormalons.
At short distances there are no
ANO strings but there still exist dynamical manifestations of the stringy 
topological condition that (external) quarks are connected by
a mathematically thin line along which the vacuum is trivial.
These manifestations may match 
ultraviolet renormalons.
    
\end{abstract}


\section{Outline of the review}

Power corrections to the parton model 
is an actual topic in QCD. Any comprehensive review of it
would ask for much more time in the oral presentation and space
in the written version than is allocated
in our case now. Moreover,
there are quite a few reviews already available 
(see, e.g., \cite{review,review1,review2,review3,review4}).
Instead we would rather like to confine ourselves to a few
background remarks and a
mini-review of some of the latest developments. 
Thus, this written version of the Talk emphasizes 
mostly power corrections associated with short-distance
non-perturbative physics, following the original papers
\cite{gpz,cnz}. 
Also, we emphasize that the classical solutions to the
Abelien Higgs model may provide a unifying dynamical framework
to reproduce the basic results of the renormalon calculus
of the power corrections.
The connection of this model to QCD is that 
the AHM realized the dual-superconductor model of the confinement 
\cite{nambu}.
The Abrikosov-Nielsen-Olesen strings
are reproducing then the generic predictions
of the infrared renormalons.
Moreover the 
short-distance behaviour of heavy quark potential potential may be closely
related to ultraviolet renormalons \cite{az3}.
Within the Abelian Higgs model the potential at short distances 
appears to posess a linear correction \cite{gpz}
which does match the ultraviolet renormalons.
These key remarks are embedded in the Talk into a general picture
of power corrections as revealed by the renormalons. 

\section{Renormalons}

Renormalons is a simple perturbative device to establish limitations 
of the perturbative approach to QCD (for reviews 
and further references see, e.g., \cite{review,review1,review3}).
In more detail, one considers an observable $O$ and
calculates it perturbatively as a series in
$\aQ$:
\be
O=O_{parton~model}\left(1+\sum_{n=1}^{\infty}a_n(\aQ)^n\right),
\label{expansion}
\ee
where $Q$ is a large mass scale characteristic for the process
considered, like total energy in
case of $e^+e^-$-annihilation into hadrons.

The expansion coefficients 
$a_n$ are calculated then in the single-renormalon-chain
approximation, i.e. in terms of 
graphs with many bubble insertions into
a gluon line. 
Then it is easy to see that the expansion coefficients grow
factorially with $n$ at large $n$:
\be
a_n~\sim~n!\left({b_0\over o}\right)^n\label{o}
\ee
where $b_0$ is the first coefficient in the expansion of the $\beta$-function,
introduced here for convenience, and $o$ is a number. This factorial 
growth of the coefficients is associated with integration over non-typical
virtual momenta $p^2$, either $p^2\ll Q^2$ (infrared renormalons) or
$p^2\gg Q^2$ (ultraviolet renormalons).

 From direct computations one finds values of the constant $o$ 
in Eq. (\ref{o}). It turns out that the exponent $o$ associated with
the UV renormalons does not depend on the variable considered:
\be
o_{UV}~=~-1
.\ee 
As for the infrared renormalons
the corresponding values of $o$ do vary from case to case.
The smallest value of $o$ known so far appears
in event shapes and, in particular, in case of the thrust:
\be
o_{thrust}~=~{1\over 2}
.\ee 
Since the series with $a_n\sim n!$ are at best asymptotical,
the perturbative QCD cannot approximate the observable $O$ to accuracy better
than
\be
\Delta~\sim~\left({\LQ^2\over Q^2}\right)^{|o|}\label{uncertainty}
.\ee
Reversing the statement, one may say that the renormalons signal 
presence of non-perturbative power corrections of order (\ref{uncertainty})
which are to be added to the perturbative series to make the answer 
for the observable $O$ unique.

Thus, single-renormalon chains provide with a unified way to determine 
for any observable the corresponding exponent
$o$.
It is important, however,
that multi-renormalon uncertainties of the perturbative expansions
(\ref{expansion}) are in fact of the same order \cite{vainshtein,review}.  
The reason is that there are two large parameters involved in
the evaluation of the renormalon-type graphs. One of them
is $n$ itself and the other is $ln(p^2/Q^2)$ where $p$ is the characteristic
virtual momentum. Multi-renormalon 
chains lose powers of the log but win
in powers of $n$. Since effectively $ln(p^2/Q^2)\sim n$ the two factors cancel
each other. As for the value of $o$, it remains 
the same.

This simple observation implies that there is no model-independent way
to relate power corrections to various observables even with the 
same exponent $o$.
Indeed, to do so one has now to evaluate any number of the renormalon
chains which is impractical.

\section{Tube model and IR renormalons.}

The sketchy presentation in the previous section
was to convince the reader that the renormalons is 
rather a kinematical framework than a dynamical one.
Indeed, even evaluating the renormalon contributions with all the
coefficients involved does not allow to judge whether the power
correction is big or small since the estimate (\ref{uncertainty})
does not fix the overall coefficient.
If the corrections are small numerically, 
on the other hand, it is very 
difficult to dig the power corrections out from the 
thick layer of pure perturbative terms.
Also, the issue of multi-renormalon contributions makes the 
problem non-tractable by direct computations.

Thus, we need models. And at first sight, no particular
mode is encoded in the values of the exponents $o$ calculable
via renormalons.

Still, one may argue that the tube model has good chances  \cite{az,review}
to match
the infrared renormalons. 
For the sake of definiteness we concentrate on the thrust $T$.
The argument starts then with an explicit evaluation of 
the one-renormalon contribution to  
$<1-T>$. Moreover, to perform the calculation one notes that the
role of the renormalon-type graphs is to replace constant $\alpha_s$ 
appearing in one-loop
by the running coupling $\ak$. Indeed, this is the function of
bubble insertions into the gluon line,
to clarify the argument of the running coupling. Thus, 
one integrates over $k_{\perp}$ of the gluon
at the last step and separates the infrared sensitive piece:
\be
<1-T>_{1/Q}\approx{2C_F\over \pi Q}\int_0^{\sim~Q}dk_{\perp}\ak\label{message}
\ee
where the integral is understood in such a way that only contribution 
of the Landau pole is kept and $2Q$ is the total CM energy.

The message brought by the Eq.(\ref{message}) is quite clear. 
Namely, we should separate
the
contribution of soft gluons with $k_{\perp}\sim\LQ$ , reserve
for an effective coupling of order unit
for the emission of such gluons, and declare this piece
to be a parameterization of non-perturbative contributions to the thrust.
The most important observation is that the leading correction is
of order $\LQ/Q$. 

To establish a connection of this derivation to the tube model 
(see, e.g., \cite{barreiro}) one proceeds as follows.
First, for two-jet events the thrust is related to (heavy- and light-)
jet masses:
\be
<1-T>_{two-jet}~=~{M_h^2+M_l^2\over 4Q^2} 
\ee
Moreover,
\begin{equation}
\left({M_{jet}^2\over Q^2}\right)_{non-pert}~
\approx{2\langle k_{\perp}\rangle\over Q}\label{bryan}
\end{equation}
where $\langle k_{\perp}\rangle\sim 0.5 GeV$ is the non-perturbative
transverse momentum of the hadrons inside the jet introduced
by the model. 

Clearly the two equations (\ref{bryan}) and (\ref{message})
refer to similar mechanisms of generating the power corrections
through introduction of an intrinsic $\langle k_{\perp}\rangle \sim \LQ$.
It is no surprise then that renormalons reproduce, for example,
the tube-model relation between the $1/Q$ corrections to the thrust and the 
$C$ parameter:
\be
\langle 1-T\rangle_{1/Q}~=~{2\over 3\pi}\langle C\rangle_{1/Q}.
\label{safe}\ee
Moreover, within the tube model one predicts:
\be
\left({\langle M_h^2\rangle\over Q^2}\right)_{1/Q}~\approx
\left({\langle M_l^2\rangle\over Q^2}\right)_{1/Q}.\label{more}
\ee
The experimental data 
\cite{fernandez} agree with both
(\ref{safe}) and (\ref{more}) although with different accuracy.
 
Derivation of the trivially-looking 
Eq. (\ref{more}) within some of the renormalon-based approaches
proved not so simple. The point is that
the standard renormalon technique which results in
Eq. (\ref{message})
implies at first sight that the renormalon contributions
are a small fraction of the ordinary perturbative graphs.
Thus, if the thrust is calculated to the order $\aQ$ then,
one could argue, non-perturbative contribution to the
jet masses can be counted also only once. 
Then the prediction would be that 
one of the jets has no nonperturbative contribution
to its mass at all and is doomed to have a strictly
vanishing mass which may not
be true experimentally of course.
Thus, one is led to assume that two-loop nonperturbative pieces
should be kept track of even if the perturbative $(\aQ)^2$ correction
is neglected. This phenomenon can be dubbed as an enhancement of
nonperturbative corrections \cite{review}. 

The prediction (\ref{more}) is natural 
within the universality picture \cite{az,kg}
which keeps terms which contribute perturbatively  
in differential distributions of the variables 
as $\alpha_s^mln^kQ$ and continues such 
terms to the infrared region
(where, in general, they do not dominate any 
longer). On the other hand, it contradicts
the large $N_f$- or naive-nonabelianization schemes
(for review and references see, e.g., \cite{review2,review3}) 
which are tailored to keep only one-renormalon chains.
The dispersive approach to the coupling (for a review and 
further references see \cite{review4})
in its original form \cite{dispersive}
also results in $M_l^2=0$. However, more recently it was
elevated to a two-loop level \cite{milan} and 
produces now predictions similar to the model of Refs. \cite{az,kg}.
It is worth emphasizing, however, that the renormalon-based frameworks
on the two-loop level
do not reduce automatically to the tube model since in higher orders
the soft gluon emission includes
non-factorizable contributions as well \cite{nason}. 

Infrared renormalons were applied also to power corrections in DIS
\cite{stein} and to shape variable distributions \cite{distribution}.
In these cases the predictions are richer since the power corrections
depend on an extra variable, like the Bjorken $x$ in case of DIS.
The renormalon-induced power corrections to 
the structure functions $F_2(x,Q^2)$ tend to
reproduce the gross features of the tube model \cite{ellis}.
In case of the structure functions $F_L(x,Q^2)$ the renormalon
predictions appear to depend on details of introducing 
the infrared cut off \cite{az5}, which is probably due to the fact that $F_L$
vanishes in the parton-model approximation.

To summarize, both the tube model with intrinsic $<k_{\perp}>\neq 0$  
and infrared renormalons predict the same values of the exponent $o$.
As for relations between power corrections to various observables,
the renormalon-based predictions do not contradict the tube model
but provide, generally speaking, with a more general and vague framework. 

\section{Ultraviolet renormalons}

As is mentioned above, the 
leading ultraviolet renormalon brings perturbative expansions of
the form: \be \left(\sum_na_n\alpha_s^n(Q^2)\right)_{UV}~\sim~ \sum_n
n!(-b_0)^n\alpha_s^n(Q^2)
\label{uv}.\ee 
Because of the sign oscillations it is quite common to apply
the Borel summation. 
The summation amounts to replacing the growing branch of
the products $|a_n\alpha_s^n|$ by its integral representation:
\be \sum_{N_{cr}}^\infty n!(-b_0)^n\alpha_s^n \rightarrow \int
{(\alpha_sb_0 t)^{N_{cr}}exp(-t)dt\over 1+\alpha_sb_0t}\label{renorm}
\ee where $N_{cr}=1/b_0\alpha_s$ is the value of $n$ for which the
absolute value of the terms in the series reaches its minimum. 
The
right-hand side is readily seen to be of order: \be
{1\over 2} \left(a_n\alpha_s^n(Q^2)\right)_{n=N_{cr}}\sim {\LQ^2\over
Q^2}.\ee It is amusing to observe that this correction comes from huge
virtual momenta of order $p^2\sim Q^2\cdot exp(N_{cr})
\sim Q^4/\LQ^2$.

Although at first sight
the Borel summation may look an arbitrary procedure 
it can be substantiated in the following way \cite{beneke}.
Instead of expanding in $\aQ$ as in Eq. (\ref{expansion})
one could expand in $\am$. Then the uncertainty of the perturbative 
expansions $\Delta$ (see Eq. (\ref{uncertainty})) does not stay invariant 
for a sign-alternating series but satisfies instead:
\be
\Delta (\am )~=~\Delta (\aQ ){Q^4\over \mu^4}\ee
and can be made therefore arbitrary small by choosing $\mu^2$ large enough.
It seems obvious, furthermore, that the series in $\am$ converges to the Borel
sum of the series in $\aQ$.

Note that accepting the Borel summation does not imply elimination
of the $\LQ^2/Q^2$ corrections due to the UV renormalons. They are still
generated through the summation procedure (\ref{renorm}).
Moreover, since the multi-renormalon contributions are not suppressed
\cite{vainshtein} there are no practical ways to evaluate
the $\LQ^2/Q^2$ terms by applying the Borel summation alone.
From this general point of 
view there is no much
difference from the case of IR renormalons. Numerically, of course,
different corrections can be of very different scales.

To probe the $\LQ^2/Q^2$ corrections of an UV origin we should consider
quantities which do not receive similar corrections from the infrared.
For example, the DIS would be a wrong choice since IR renormalons
already induce $\LQ^2/Q^2$ corrections. Thus, the central object
to study UV renormalons \cite{yamawaki}
are the vacuum correlators of the currents $j$
with various quantum numbers:
\begin{equation}
\Pi_j (Q^2)~=~i\int exp(iqx)\langle 0|T\{j(x),j(0)\}|0\rangle ,\label{corr}
\end{equation}
where $q^2\equiv -Q^2$ and we suppressed the Lorentz indices and
assumed that the currents are normalized to have zero anomalous dimension.
Moreover, to get rid of the UV divergences
one studies usually $\Pi (M^2)$ where \cite{svz}
\be
\Pi_j (M^2)~\equiv~{ Q^{2n}\over (n-1)!}\left({-d\over dQ^2}\right)^n\Pi_j (Q^2)
\ee
in the limit where both $Q^2$ and $n$ tend to infinity so that their
ratio $M^2\equiv Q^2/n$ remains finite. 
According to the standard OPE:
\bea
\Pi_j(M^2)\approx (parton~model)\cdot\\ \nonumber \cdot
\left(1+{a_j\over lnM^2/\LQ^2}+{c_j\over M^4}+O((lnM^2)^{-2}M^{-6})\right)
\label{salient}\eea
where the constants $a_j,c_j$ depend on the channel, i.e. on the quantum
numbers of the current $j$. The terms of order $1/lnM^2$ and $M^{-4}$
are associated with the first perturbative correction and the gluon
condensate, respectively.

A salient feature of Eq. (\ref{salient})
is the absence of $\LQ^2/Q^2$ terms which is due to the fact that there 
are no gauge invariant operators of dimension d=2 which could have
vacuum-to-vacuum matrix elements.
Now, accounting for the UV renormalons brings $\LQ^2/Q^2$ 
terms which go beyond the standard OPE:
\bea
\Pi_j(M^2)\approx (parton~model)\cdot\\ \nonumber \cdot
\left(1+{a_j\over lnM^2/\LQ^2}+{b_j\over M^2}+{c_j\over M^4}+...\right).
\label{correl}\eea
In section 7 we review briefly a model to evaluate the
coefficients $b_j$ in various channels.

Another physical quantity which can be used to isolate the
$\LQ^2/Q^2$ corrections of the UV origin 
is the heavy quark potential $V(r)$ at short distances \cite{az3}.
Indeed, it is obvious from dimensional considerations that
the leading power correction to the perturbative potential  
at short
distances $r$
is now linear in $r$:
\be
\lim_{r\rightarrow 0}V(r)=-{4\alpha_s(r)\over 3r}+\sigma r\label{pot}
~~~~~(non-perturbative~ 1/Q^2~corrections)\label{nonstandard}.\ee
On the other hand, within the so to say standard QCD the leading 
power correction at {\it short} distances is of order $r^2$:
\be
\lim_{r\rightarrow 0}V(r)=-{4\alpha_s(r)\over 3r}+cr^2
~~~~~(standard~ QCD)\label{standard}
.\ee
 This 
conclusion is 
based solely on the assumption that the nonperturbative 
fluctuations in QCD are of large
scale, $\sim \LQ^{-1}$ (for references and further explanations see
\cite{az3}). Thus, the introduction of 
the linear correction to the potential through
the new $\LQ^2/Q^2$ terms from the ultraviolet
assumes small-size nonperturbative field configurations.
A particular picture of the vacuum properties which results in
this effect is described in Ref \cite{az3}.

\section{ANO strings.}

We will consider now 
the Abelian Higgs model (AHM) describing interactions of 
a $U(1)$ gauge field $A_{\mu}$ with a charged scalar field $\Phi$
in the Higgs phase when the scalar field condenses. The model is
defined by its action :
\be\label{AHM_action}
S= \int d^4x \left\{
\frac{1}{4e^2} F^2_{\mu\nu} + \frac{1}{2} |(\diff - i A)\Phi|^2 + 
\frac{1}{4} \lambda (|\Phi|^2-\eta^2)^2
\right\}\label{ahm}
\ee
where $F_{\mu\nu}\equiv\diff_{\mu}A_{\nu}-\diff_{\nu}A_{\mu}$ and the 
complex scalar field $\Phi$ carries electric charge. The physical 
vector and scalar particles are massive, $m^2_V=e^2\eta^2, m_H^2=
2\lambda \eta^2$ and for the sake of definiteness we assume
$m_H>m_V$.

The model (\ref{AHM_action}) is famous 
to provide with a relativistic analog of
the Landau-Ginsburg theory of superconductivity. 
Namely,
if one introduces a \ma~ pair as an external 
probe its static potential
$V(r)$ grows linearly with the distance $r$ at large $r$: 
\be
lim_{r\rightarrow \infty}V(r)~=~ kr.\label{growth}
\ee
This property is crucial for the dual-superconductor model
of confinement \cite{nambu}. Within this 
picture, one thinks about an external $\bar{Q}Q$ pair 
in the environment of a condensed scalar field which carries 
a non-zero monopole charge. Up to the change 
of the notations to the dual ones,
the physics is the same as described by (\ref{ahm})
and for the sake of definiteness we shall not change the notations,
i.e. will be considering interaction of a \ma~ pair within the AHM.

The relevance of the AHM to QCD is most obvious
in the $U(1)$ projection of QCD \cite{thooft}
when the diagonal gluons are treated as $U(1)$, i.e. photonic, 
gauge fields
while the other (non-abeliean) 
components play the role of charged vector fields.
Moreover, 
there exist
detailed numerical simulations on the lattice which confirm 
the dual-superconductor picture of the confinement
(for review and references see \cite{polikarpov}).

The growth of the potential at large distances, i.e. at 
$r\gg\LQ^{-1}$
is well understood in terms of formation of the Abrikosov-Nielsen-Olesen 
(ANO) strings
(for review and further references see, e.g., \cite{baker}). The ANO strings 
\cite{ano}
are solutions to the classical equations of motion corresponding to the action (\ref{ahm}) with cylindrical symmetry
and carrying a (quantized) magnetic flux $\Phi=2\pi/e$.
The magnetic field is distributed within the string as:
\be
|H|={m_V^2\over e}K_0(m_Vr)\rightarrow_{r\rightarrow\infty}
\eta\sqrt{{\pi\eta\over 2e r}}e^{-m_Vr}
\ee
where $r$ is the distance in the transverse direction.
As for the scalar, or Higgs field it disappears on the axis,
\be
\lim_{r\to 0}|\Phi (r)|~=~0
\ee 
and approaches its vacuum value at large $r$:
\be
\lim_{r\to \infty}|\Phi (r)|~=~\eta+(const)~e^{-m_Hr}.
\ee
A string may either be closed or end up with monopoles.
In the latter case the const $k$ in Eq. (\ref{growth})
is the energy density per unit length of the ANO string,
\be
k~\sim~{\pi\eta^2\over 2}ln{m_H^2\over m_V^2}
\ee
where we assumed that $m^2_H\gg m_V^2$. If the latter inequality
is not satisfied, the log is replaced by a smooth function of the ratio 
$m_H/m_V$.
In case of QCD the formation of the ANO string
is confirmed by the numerical simulations in the $U(1)$ projection
and the space structure of the fields inside the string
is indeed well described by the classical equations of motion \cite{bali1}.

After this brief description of the ANO string
and its relevance to QCD, we can go back
to the power correction.
The relation to the tube model is rather self-evident. Indeed,
the ANO string introduces a characteristic transverse scale,
$x_{\perp}\sim m^{-1}_V\sim (\LQ)^{-1}$.
In the momentum space, we have $k_{\perp}\sim m_V$ 
and, as is discussed in the preceding sections, this is what is needed
to match the power corrections predicted by the renormalons. 
Numerically \cite{bali1}, 
the value of $m_V$ is quite large, $m_V\sim 1 GeV$
so that the corresponding power corrections are large, may be even too large\
to match the phenomenological estimates.

\section{Heavy quark potential at short distances. Dirac strings.}

Let us now discuss short-distance physics,
looking for a possible link to the UV renormalons.
At first sight, there is nothing left from the ANO string if we go to
short distances, $r\ll m_V^{-1}$.
And this is of course true.
However, a closer analysis reveals that there exist in fact 
two other stringy objects which can survive at short distances 
\cite{gpz,gpz1}.
Indeed, in the so called London limit, $m^2_H\gg m^2_V$, or
${\lambda}\to \infty$ with $\eta$ fixed, the size of the
Higgs core of the ANO string
is much smaller than the size of the magnetic field and one may
consider the distances $m_V^{-1}\gg r\gg m_H^{-1}$. 
Then the question is, what happens to the linear piece 
(\ref{growth}) in the potential at such distances.

Even more intriguingly, there is an object in the theory which
is nothing else but a mathematically thin line.  This is the line
connecting the monopole and anti-monopole
singled out through vanishing of the scalar field $\Phi$ along this line.
The existence of such a line connecting 
the monopoles can be considered
as a topological condition: the boundary of the world sheet with
$\Phi\equiv 0$ can be nothing else but world-trajectories
of monopoles \cite{thooft}.
However, the language of the Dirac string \cite{dirac} may provide
with a better insight. The existence of the Dirac string
follows, as usual, from the magnetic flux conservation.
 
The possibility of a dynamical manifestation of the Dirac string
stems from the fact that it cannot coexist
with $\Phi\neq 0$ so that $\Phi$ vanishes along the string.
Indeed, the self-energy of the Dirac string,
is normalized to be zero in the perturbative 
vacuum. To justify this assumption one can invoke duality and 
ask for equality of the self-energies of electric and magnetic
charges. However, if the Dirac string would be embedded into a vacuum
with $<\Phi>\neq 0$ then its energy would jump to infinity because
there is a term $\sim\Phi^2A_{\mu}^2$ in the Hamiltonian and 
$A_{\mu}^2\rightarrow \infty$ for a mathematically thin Dirac string.
Hence, $\Phi=0$ along the string.
One may say that the Dirac strings always rest on the perturbative vacuum
which is defined as a vacuum state obeying the duality principle. 
Therefore, even in the limit $r\to 0$ there is a deep well in
the profile of the Higgs field $\Phi$. This might cost energy which 
is linear with $r$ even at small $r$.

Consider first the London limit. It
can be studied
analytically \cite{gpz1}. 
The result is that the answer for the potential is the same
as if we had an ANO string fully developed \cite{gpz1}:
\bea
V(r)=-{\pi\over e^2}{e^{-m_Vr}\over r}+\\ \nonumber
+{\pi m_V^2\over 2e^2}\left(
rln{m_H^2\over m_V^2}-{2\over m_V}+\int_0^{\infty}dk_{\perp}^2
{e^{-r\sqrt{k_{\perp}^2+m_V^2}}\over
(k_{\perp}^2+m_V^2)^{3/2}}\right).
\label{londonlimit}
\eea
Note that the linear correction to the potential
persists despite of the fact that the distances considered are much
smaller that the transverse size of the magnetic field 
within the ANO string, Naively, one would expect large edge effects
which turn  not to be there, however. One can consider this as an example
of a dual description. Namely, in one language the ANO string is bulky,
$x_{\perp}\sim m_V^{-1}$ while in the dual description it is 
entirely characterized
by position of its center, $\Phi =0$.

If we make the next step and consider $r\ll m_V^{-1},m_H^{-1}$
then nothing is left from the original ANO strings. The field configuration
is close to that of a magnetic dipole. Nevertheless, we 
still have a Dirac string (see above)
which is manifested through a boundary condition 
$\Phi =0$ along the straight
line connecting the monopoles 
imposed on the solutions of the classical equations of motion.
The potential at short distances is dominated by a Coulomb-like
contribution. At intermediate distances the potential can be still
found by solving the equations of motion numerically \cite{numerical}.

However, until very recently \cite{gpz} there were no dedicated
studies of the power correction to the Coulomb-like potential
at short distances.
The result of this recent study is that the linear correction
to the potential still persists at $r\to 0$ (see Eq. (\ref{pot})).
Moreover, for 
$m_V\approx m_H$ the result for the linear correction at short
distances
is especially simple:
\be
\sigma~\approx~k\label{approx}
\ee
where 
$\sigma r$ gives the linear correction at {\it short} distances
(see Eq. (\ref{pot})) while $kr$ refers to the linear potential 
at {\it large} distances,
see Eq. (\ref{growth}).
Note that Eq. (\ref{approx}) approximately holds in spite of a complete
change of the dynamical mechanism for the linear piece of the potential,
which is the change from the ANO to Dirac strings and is true only for
$m_V\approx m_H$.
  
It is amusing that the numerical simulations of the real QCD
also give $k\approx \sigma$ \cite{bali2} and (separately)
$m_V\approx m_H$ \cite{bali1} down to the distances studied so far. 
Moreover, the fact the slope $k$ is not changing 
somehow masks the fact that the linear term $\sigma r$ at short
distances is in contradiction with the standard QCD,
see Eq. (\ref{standard}).
Thus, if no new short-distance 
non-perturbative effects are introduced the persistence of the
relation (\ref{approx}) in the lattice measurements is
to be attributed either to not-yet-small-enough distances $r$  
(the measurements extend down to $r\simeq 0.1 fm$, though) or to large
error bars which do could have confused the fact that
the linear piece actually vanishes.
 
On the other hand, a non-vanishing linear correction to the
potential at $r\to 0$ is predicted within the classical 
approximation to the
Abelian Higgs model \cite{gpz}.
Moreover, it 
is a direct dynamical manifestation of the Dirac string,
or of the topological condition that the vacuum 
is trivial along a line connecting the monopoles  \cite{gpz}.
If the AHM is to imitate QCD, this is for the first time 
that there a  dynamical mechanism proposed which incorporates
naturally the results of the measurements \cite{bali2}
of the potential on the lattice.
Since in the $U(1)$ projection of QCD \cite{thooft}
the monopoles appear as singular, or point-like objects \cite{polikarpov1}
there are good chances, to my mind, that the $U(1)$ description 
outlined above is a valid approximation to short distances.
This question has not been studied in any detail so far, however.

Coming back to the renormalons,
the linear correction to the $\bar{Q}Q$ potential at short distances
 would perfectly
match, in the language of the preceding sections, the UV renormalons.

\section{Short-distance tachyonic gluon mass.}

The linear correction to the potential at short distances discussed
in the previous section gives an example of a $1/Q^2$ 
correction of UV origin. 
It would be interesting to relate this correction to the $1/M^2$ 
terms
in the current correlators (\ref{correl}).
At this moment, however, a small-size-string correction
to the current correlators
cannot be evaluated from first principles.  
Instead  we will review briefly in this section a simplified 
phenomenology in terms of a tachyonic gluon mass which is assumed to mimic
the short-distance nonperturbative effects \cite{cnz}.

To motivate this, rather drastic assumption we simply notice that
the linear term in the potential at short distances  (see Eq. (\ref{pot}))
can be imitated \cite{vz2} by the Yukawa 
potential with a gluon mass $\lambda $:
\be
{4\alpha_s\over 6}\lambda^2~\sim~-~\sigma\label{mass}
.\ee
The tachyonic sign for the $\lambda^2$ arises because of
the positive string tension of the small-size strings
found in the preceding section.
This notion of the short-distance gluon mass  
can be consistently used at one-loop
level as well 
without running into a conflict with the gauge invariance.
This observation is crucial to extend the phenomenology from
the short-distance potential to the correlators (\ref{correl}).

Concentrating therefore on the phenomenology of these correlators
we notice first that the basic quantity here is the scale $M^2_{crit}$
at which the parton model gets violated by the power corrections in
various channels \cite{novikov}. More specifically,
$M^2_{crit}$ is defined as the value of $M^2$ at which 
the power corrections become, say, 10\% from the unit. The meaning of 
$M^2_{crit}$ is that at lower $M^2$ the power corrections blow up.

In the $\rho$-channel, $M^2_{crit}$ is controlled by the $M^{-4}$ term
and: 
\be
M^2_{crit}(\rho-channel)~\sim~0.6~GeV^2\label{normal}
\ee
where the numerical value of $M^2_{crit}$ is fixed by the
magnitude of the gluon condensate
$<\alpha_s(G_{\mu\nu}^a)^2>$. Moreover, it agrees well with independent 
evaluation of $M^2_{crit}$ from the experimental data
on the $e^+e^-$ annihilation \cite{svz}. 

If one proceeds to other channels, in particular to the 
$\pi$-channel and to the $0^{\pm}$-gluonium channels, nothing special
happens to $M^2_{crit}$ associated with the gluon condensate.
However, it was determined from independent arguments that the
actual values of $M^2_{crit}$ do vary considerably in these
channels \cite{novikov}:
\be 
M^2_{crit}(\pi-channel)~\ge~ 1.8~GeV^2\label{pion}\ee
\be
M^2_{crit}(0^{\pm}-gluonium~channel)~\ge~ 15~GeV^2\label{gluonium}
.\ee
These lower bounds on $M^2_{crit}$ are obtained from the values of
$f_{\pi}$ and of the quark masses in the pion channel, and from a low-energy 
theorem in the gluonic channel. Such values of $M^2_{crit}$ 
cannot be reconciled with the assumption that the gluon condensates
controls $M^2_{crit}$ in all the channels.

Now, that we have introduced a short-distance gluon tachyonic mass,
the coefficients
$b_j$ are calculable in terms of $\lambda^2$ \cite{cnz}:
\be
b_{\pi}\approx 4b_{\rho}={4\alpha_s\over 3\pi}c_{gluonium}=
-{4\alpha_s\over \pi}\lambda^2.
\ee
Phenomenologically,
in the $\rho$-channel there are severe restrictions \cite{narison}
on the new term $b_{\rho}/M^2$:
\be
b_{\rho}~\approx~(0.03-.07)~GeV^2\label{constr}
.\ee
Remarkably enough, the sign of $b_{\rho}$
does correspond to a tachyonic gluon mass.
Moreover, when interpreted in terms of $\lambda^2$ 
the constraint (\ref
{constr}) does allow for a large $\lambda^2$, say, $\lambda^2=-0.5GeV^2$.

Another crucial test is the effect of $\lambda^2\neq 0$ on the 
value of $\alpha_s(M_{\tau}^2)$ as determined from the width $R_{\tau}$
of the leptonic $\tau$-decays. Indeed, the running coupling is well known 
independently. It turns out that $\lambda^2\approx 0.5 GeV^2$
brings $\alpha_s(M_{\tau}^2)$ down by about 10\%.
The sign of the change is again the right one to improve the agreement
with $\alpha_s(m_Z^2)$ and the absolute value of the change
is within experimental uncertainties.

As for for the $\pi$-channel one finds now a new value of $M^2_{crit}$
associated with $\lambda^2\neq 0$:
\be
M^2_{crit}(\pi-channel)\approx~4\cdot M^2_{crit}(\rho-channel)\ee
which fits nicely the Eqs.
(\ref{normal}) and (\ref{pion}) above.
Moreover, the sign of the correction in the $\pi$-channel 
is what is needed for phenomenology \cite{novikov}.
This can be considered as another crucial test of the
tachyonic sign of the $\lambda^2$.
Fixing the value of $c_{\pi}$ to bring 
the theoretical $\Pi_{\pi}(M^2)$
into agreement with the phenomenological input one gets
\be
\lambda^2~\approx~-0.5~ GeV^2
.\ee

Finally, we can determine the new value of $M_{crit}^2$ in the 
scalar-gluonium channel and it turns to be what is 
needed for the phenomenology,
see Eq (\ref{gluonium}).
Thus, qualitatively the phenomenology with a tachyonic gluon mass
which is quite large numerically stands well to a few highly nontrivial
tests.

It is worth emphasizing that the $\lambda^2$ terms represent 
nonperturbative physics and limit in this sense the range of applicability
of pure perturbative calculations. 
This nonperturbative piece may well be 
much larger than some of the perturbative corrections which are 
calculable and calculated nowadays.

Further crucial tests of the model with the tachyonic gluon mass could be 
furnished with measurements of various correlators $\Pi_j(M^2)$
on the lattice \cite{cnz}.

\section{Conclusions}

We discussed briefly the status of the power corrections
associated both with IR and UV regions. 
We argued that
the Abrikosov-Nielsen-Olesen and Dirac strings of the $U(1)$ projection of QCD
appear to
match renormalons, infrared and ultraviolet respectively. 
Phenomenologically, there is room for a new, relatively large 
$\LQ^2/Q^2$ correction coming from the ultraviolet.
If confirmed, such a correction would be of great interest. 

\section{Acknowledgements}

This review is based to a great extent on the original papers
\cite{gpz,gpz1,cnz} and I am gratefully acknowledging 
collaborations with
K.G. Chetyrkin, F.V. Gubarev,
S. Narison and M.I. Polikarpov.
I am also thankful to R. Akhoury, 
V.A. Rubakov, L. Stodolsky, A.I. Vainshtein, A. Zhitnitsky  
for interesting discussions and remarks.

I am grateful to the Organizing Committee of the RADCOR '98
(Barcelona, September '98) and especially to Prof. J. Sola for the
invitation and hospitality.

\end{document}